\documentclass{aastex631}

\usepackage{amssymb}
\usepackage{amsmath}
\usepackage{cases}

\shorttitle{Flare turbulence}
\shortauthors{Ruan et al.}
\graphicspath{{./}{figures/}}

\begin{document}

\title{MHD turbulence formation in solar flares: 3D simulation and synthetic observations}

\author[0000-0001-5045-827X]{W. Ruan}
\affiliation{Centre for mathematical Plasma Astrophysics, Department of Mathematics, KU Leuven, \\
Celestijnenlaan 200B, B-3001 Leuven, Belgium}
\email{wenzhi.ruan@kuleuven.be}

\author{L. Yan}
\affiliation{Key Laboratory of Earth and Planetary Physics, Institute of Geology and Geophysics, 
		 Chinese Academy of Sciences}

\author{R. Keppens}
\affiliation{Centre for mathematical Plasma Astrophysics, Department of Mathematics, KU Leuven, \\
Celestijnenlaan 200B, B-3001 Leuven, Belgium}

\accepted{Oct 17, 2022}
\submitjournal{ApJ}

\begin{abstract}
Turbulent plasma motion is common in the universe, and invoked in solar flares to drive effective acceleration leading to high energy electrons. 
Unresolved mass motions are frequently detected in flares from extreme ultraviolet (EUV) observations, which are often regarded as turbulence.
However, how this plasma turbulence forms during the flare is still largely a mystery.
Here we successfully reproduce observed turbulence in our 3D magnetohydrodynamic simulation where the magnetic reconnection process is included. 
The turbulence forms as a result of an intricate non-linear interaction between the reconnection outflows and the magnetic arcades below the reconnection site, in which the shear-flow driven Kelvin-Helmholtz Instability (KHI) plays a key role for generating turbulent vortices.
The turbulence is produced above high density flare loops, and then propagates to chromospheric footpoints along the magnetic field as Alfv\'enic perturbations.
High turbulent velocities above 200 km s$^{-1}$ can be found around the termination shock, while the low atmosphere reaches turbulent velocities of 10 km s$^{-1}$ at a layer where the number density is about 10$^{11}$ cm$^{-3}$. 
The turbulent region with maximum non-thermal velocity coincides with the region where the observed high-energy electrons are concentrated, demonstrating the potential role of turbulence in acceleration.
Synthetic views in EUV and fitted Hinode-EIS spectra show excellent agreement with observational results.
An energy analysis demonstrates that more than 10\% of the reconnection downflow kinetic energy can be converted to turbulent energy via KHI.
\end{abstract}

\section{Introduction} \label{sec:intro}

Solar flares actively drive space weather, which affects interplanetary space and the Earth environment and atmosphere through energetic particles, strong EUV/X-ray emissions and associated coronal mass ejections (CMEs). In a solar flare event, up to 10$^{33}$ erg of free energy stored in the solar magnetic field can be released via the magnetic reconnection process, and up to half of this energy is used to produce energetic electrons \citep{Aulanier2013,Aschwanden2017}. 
Magnetic reconnection can develop at current sheets, marking locations where a component of the magnetic field reverses \citep{Parker1963}.
This changes the configuration of magnetic field lines, converts magnetic energy to thermal and kinetic plasma energy and produces fast flows that leave the reconnection site along the current sheet (a recent review see \citet{Yamada2010}).
In a typical flare event, reconnection is generally believed to occur at coronal heights, and the downward reconnection outflows lead to -- and interact with -- magnetic arcades in lower regions of the corona. 
These arcades are filled with hot and dense ionized gas and they become bright in the images at soft X-rays (SXR) and specific EUV passbands \citep{Shibata1995}.
Information on energetic electrons produced in solar flares is often derived from hard X-rays (HXR, photons with energy $> 20$ keV), as energetic photons are believed to be produced by electrons via the bremsstrahlung mechanism (e.g. \citealp{Kontar2011}). 
Strong and isolated HXR sources are often observed (1) at the footpoints of the hot and dense arcade flare loops at chromospheric heights (0.5 - 3 Mm above the photosphere) and (2) near the top of the flare loops (e.g. \citealp{Masuda1994,Tomczak2007,Su2013}).
Hence, a large amount of energetic electrons of energy $> 20$ keV must be produced during these explosive events  \citep{Hudson1995,Tomczak2001,Krucker2008}.

In the study of solar flares, promising mechanisms to produce energetic electrons are turbulence acceleration, as well as shock acceleration and direct current (DC) electric field acceleration \citep{Aschwanden2005}. 
For the former mechanism, unresolved mass motions with velocities exceeding 100 km s$^{-1}$ have been discovered in flares for decades (e.g. \citealp{Doschek1980,Gabriel1981,Antonucci1982}). The unresolved mass motions are often thought of as turbulence, although it is impossible to know whether typical energy cascades are happening from remote observations.
Recent observations with higher resolution data demonstrate that this plasma turbulence is not really localized, and covers the entire flare region including flare loop top, legs, footpoints and the region above the looptop (e.g. \citealp{Doschek2014,Kontar2017,Jeffrey2018,Stores2021}). Peaks in turbulence velocity values tend to show up above the high density flare loops, reaching 100-200 km s$^{-1}$, while footpoints have lower turbulent velocities of a few tens km s$^{-1}$. 
Turbulence has been frequently reproduced in 2D simulations, but the obtained turbulence is more localized (e.g. \citealp{Fang2016,Ruan2018,Ye2019,Wang2022}). \citet{Wang2022} invoke the Kelvin-Helmholtz instability in their 2D settings, to explain the observed wiggling of the current sheet above.
In contributions to solar flare research, we investigated \citep{Ruan2020} the interplay between 2D magnetohydrodynamic flare evolutions as coupled to analytic prescriptions of the accelerated electron beams, where we could reproduce both HXR source regions; and evolve the flare far into the postflare regime to reproduce flare-driven coronal rain~\citep{Ruan2021}. Recently, a fully 3D solar flare simulation shows that turbulence can be produced in flares, with a clear role played by a mixture of the Rayleigh-Taylor (RTI) and the Richtmyer-Meshkov instability (RMI) at the interface between the reconnection termination shock and the flare arcade, leading to finger-like supra-arcade downflows \citep{Shen2022}. Our present 3D simulation will augment this result, by looking especially at the spatial distribution and the values of the non-thermal velocities obtained, and by making detailed comparisons to observations. Our present 3D simulation goes one step further to investigate the cause and consequence of the plasma turbulence in the entire flare region by successfully reproducing the observational features of the plasma turbulence, including the spatial distribution and characteristic velocities.

\section{Method} \label{sec:method}

The simulation is performed with the open-source MPI-AMRVAC code \citep{Xia2018,Keppens2021}. The square simulation box has a domain of -50 Mm $\leq x \leq$ 50 Mm, -50 Mm $\leq y \leq$ 50 Mm and 0 $\leq z \leq$ 100 Mm. The minimal resolution is $64 \times 64 \times 64$, but an equivalent high resolution of $1024 \times 1024 \times 1024$ is achieved via employing 5 levels in our adaptive mesh refinement strategy, making our smallest cells less than 100 km across (the \cite{Shen2022} grid cell size was at 260 km). Gravity, thermal conduction and optically thin radiative losses are included, where the cooling curve comes from \citet{Colgan2008}. A spatially varying, but temporally invariant background heating is employed to balance the radiative loss and maintain a corona. The governing equations are identical to \citet{Ruan2020}. A magnetic-field-line-based transition region adaptive conduction (TRAC) method is adopted to ensure that underresolving the sharp transition region variations -- which would need better than 30 km resolution \citep{Johnston2020,Zhou2021} -- does not lead to erroneous coronal temperature and density evolutions. 

The initial conditions for number density, temperature and background heating are similar to our 2.5D flare simulation presented in \citet{Ruan2021}, where all initial profiles are functions of height $z$ only. The initial coronal region has an electron density of order $\sim 10^9$ cm$^{-3}$ and temperature of $2-3$ MK. 
The number density and temperature profiles are obtained from a relaxation in which the model C7 temperature profile in \citet{Avrett2008} and a density profile calculated based on hydrostatic equilibrium are employed as initial conditions, where the number density at the bottom boundary is 3.7 $\times$ 10$^{14}$ cm$^{-3}$.
The initial conditions for magnetic field are modified from those in \citet{Ruan2021}, which is given by
\begin{eqnarray}
B_x &=& \sqrt{B_0^2-B_z^2}, \\
B_y &=& 0, \\
B_z &=&
\begin{cases}
-B_0, & \quad y<-\lambda \\
B_0, & \quad y>\lambda \\
B_0 \sin [\pi y/(2\lambda)], & \quad else \\
\end{cases}
\end{eqnarray}
where $B_0=30$ G is the initial magnetic field strength and $\lambda=10$ Mm.
We dissipate the thick current sheet into a thin current sheet in the pre-flare phase by adopting a resistivity given by
\begin{equation}
\eta (y,z) = \eta_1  \exp(- y^2/w_{\eta y}^2) \exp[- (z - h_{\eta1})^2/w_{\eta z}^2] \,,
\end{equation}
where $\eta_1=10^{-2}$, $w_{\eta y}= 10 \ \rm Mm$, $h_{\eta 1}=30 \ \rm Mm$ and $w_{\eta z}= 15 \ \rm Mm$. Magnetic arcades are produced in the lower atmosphere during the slow dissipation of the thick current sheet, without generating strong chromospheric evaporation. Hot and dense flare loops will only be generated above the magnetic arcades in the upcoming impulsive phase, that leads to formation of clear bright loops in the high temperature EUV images (e.g. 131\,{\AA}). The anomalous resistivity prescription is changed when a thin current sheet is formed at $t=450$ s. The new resistivity adopted then is given by
\begin{equation}
\eta (x,y,z) = \\
\begin{cases}
\eta_2 + \eta_3 \exp \{ - [x^2 + y^2 + (z-h_{\eta 2})^2]/r_{\eta}^2 \}, & 480\ \mathrm{s} \geq t > 450\ \mathrm{s}  \\
\eta_2 , & \quad t>480 \ \mathrm{s}
\end{cases}
\end{equation}
where $\eta_2=10^{-3}$, $\eta_3=10^{-1}$, $h_{\eta 2}=50 \ \rm Mm$ and $r_{\eta}= 5 \ \rm Mm$. Fast reconnection is triggered by this localized strong resistivity at the center of our thinned current sheet.
The magnetic field inside the initial current sheet contains an $x$ component, therefore the earliest formed magnetic arcades (mainly located inside/below the dense flare loop) also contain an $x$ component, and hence have a shear angle to the magnetic neutral line.
The magnetic arcades generated at the impulsive phase are almost in $y$-$z$ planes, as the $x$ component of the magnetic field inside the reconnection current sheet has been transported to the lower atmosphere and the region above the simulation box. 

Periodic boundaries are employed at the $x$ boundaries. Symmetric boundary conditions are adopted for density, pressure and $x$/$z$ components of velocity/magnetic field at the $y$ boundaries, while anti-symmetric conditions are used for $y$-components of velocity and magnetic field. Fixed boundary conditions are employed at the bottom boundary. We employ zero gradient extrapolation at the (two-layer) ghost cells of the upper boundary for density, velocity and magnetic field. 
The temperature at the top ghost cells is forced to decrease at a rate $dT/dz=-T_b/$(50 Mm) to avoid runaway high temperatures at the boundary, where $T_b$ is the instantaneous local temperature at the boundary.
A combination of the `HLL' (initials of authors Harten, Lax and van Leer in \citet{Harten1983}) approximate Riemann solver and `Cada3' (first author from \citet{Cada2009}) flux limiter is employed at the low level ($\leq 3$) grids, as the third-order limiter `Cada3' achieves high accuracy at low spatial resolution. Another combination, `HLL' with `Vanleer' (from \citet{vanLeer1974}) limiter, is adopted at high level grids located at the low atmosphere and flaring regions, as the second-order `Vanleer' limiter has better performance in high gradient regions. A strong stability preserving, three-step Runge-Kutta method is employed in time integration \citep{Ruuth2006}.

Contribution functions from the CHIANTI atomic database have been used in synthesizing EUV emissions \citep{DelZanna2015}. 
The spatial resolution of the synthesized EUV images is the same as given by the observation (using a pixel size of {435 km} for image at 131 {\AA} passband).
The synthesized 255 {\AA} spectra are assumed to have a slit width of 1450 km and a pixel resolution of 725 km $\times$ 22 m{\AA}. 
The 131 {\AA} line is mainly related to emission by Fe VIII and Fe XXI ions with peak formation temperatures of 10$^{5.6}$ K and 10$^{7.0}$ K, and the 255 {\AA} line is mainly associated with Fe XXIV ions with a peak formation temperature of 10$^{7.2}$ K \citep{Culhane2007,Lemen2012}.
Scattering effects by the instrument have been included via multiplying by a Point Spread Function (PSF), where we assume the PSF is a Gaussian function of standard deviation of 1 pixel \citep{Grigis2013}. Note that only the contribution of hot plasma ($>3$ MK) is considered in synthesizing 131 {\AA} emission, to avoid unrealistic strong emission at the low atmosphere due to underresolved transition region physics. The GOES SXR flux is calculated with the method given in \citet{Pinto2015}.

\section{Results} \label{sec:result}

\subsection{Bright EUV arcades evolution in flaring region}

\begin{figure}[htbp]
\begin{center}
\includegraphics[width=0.8\linewidth]{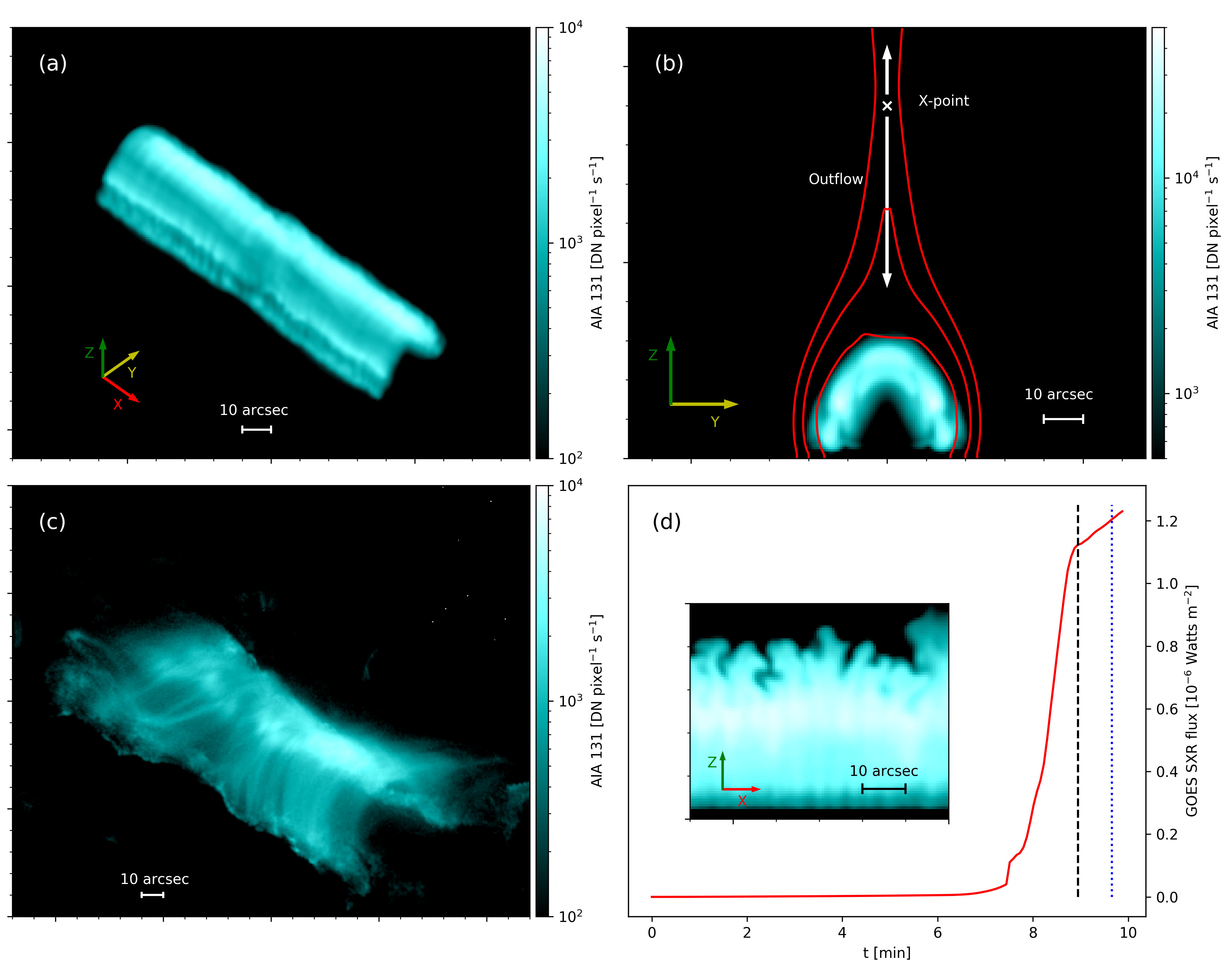}
\caption{(a) \& (b): Synthesized 131 {\AA} views of the flare loop systems obtained in our simulation. The two views are synthesized with different line of sight (LOS) directions, where the corresponding orientation axes are given. The red curves in (b) are projections of several magnetic field lines. The approximate locations of the magnetic reconnection X-point (white cross) and the reconnection outflows (white arrows) are also given in (b). Panel (c) shows a 131 {\AA} image of a flare event from Dec 26, 2011, where the LOS direction is similar to that in (a), as reported by \citet{Cheng2016}. Panel (d) gives the time development of synthesized GOES SXR flux at 1-8 {\AA} passband, where the black dashed line gives the corresponding time of the synthesized views. A cross-section in the $X-Z$ plane, in a zoomed-in view on the top of the arcade, shown in panel (d), confirms the RTI/RMI process from \citet{Shen2022}, happening at a later time referring to the blue dotted line in this panel. An animation of this figure is available. Panel (a)-(c) of the animation give synthesized 131 {\AA} views obtained with different LOS, where the corresponding orientation axes are given. Panel (d) of the animation shows the synthesized GOES SXR flux at 1-8 {\AA}, the same as panel (d) of this figure. The animation covers $\sim 10$ minutes of physical time starting at $t=0$ (real-time duration 14 s). }
\label{fig1}
\end{center}
\end{figure}

Our simulation starts with a single vertical ($x-z$ oriented) current sheet all along $y=0$, which runs through the chromosphere and the corona. Magnetic reconnection happens inside this current sheet at coronal heights, leading to formation of an extended coronal magnetic arcade system below the reconnection site. Some magnetic arcades are filled with hot ($\sim 10$ MK) and dense plasma ($\sim 10^{10}$ cm$^{-3}$) by downward flows from the reconnection site and (thermal conduction driven) upward flows from the chromosphere, respectively. This then forms high density flare loops, which are bright in images at the EUV 131 {\AA} passband. Photons at this passband are mainly released by Fe XXI ions at temperatures $\sim$10 MK in flare events. Figure~\ref{fig1} gives synthesized images at 131 {\AA} passband of the simulated flare loops, where for comparison, an actual flare event observed by the \textit{Atmospheric Imaging Assembly} (AIA) onboard spacecraft \textit{Solar Dynamics Observatory} (SDO) is also given \citep{Lemen2012}. 
The axes orientation is shown in both top panels, where $Z$ is the vertical direction in our simulation, the bottom of the chromosphere being located at $Z=0$. 
We indicated a reference bar of 10 arcsec length, to better compare with observations, noting that 1 arcsec $=$ 725 km. 
Here we use capital letters ($X$, $Y$, $Z$) to indicate directions when the length unit is arcsec and use lower case letters ($x$, $y$, $z$) when the length unit is Mm.
The hot flare loops start to form at $t\approx 7$ min due to the triggering of fast reconnection, and then the flare enters an impulsive phase. Here we focus on the generation and development of turbulence in the impulsive phase, at the time after the formation of the flare loops. Most of the figures in our paper use the data at $t\approx 9$ min (shown with a black dashed line in Fig.~\ref{fig1}d), when finger-like structures have not yet appeared. The wide initial current sheet is slowly dissipated by resistivity and then becomes a thin current sheet in the period $t < 7$ min. Magnetic arcades are also formed in this period, but without the generation of strong evaporations. Consequently, there are no bright 131 {\AA} loops or strong SXR emission during this period.

The finger-like structures, which are caused by RTI/RMI according to \citet{Shen2022}, also appear in our simulation, but clearly at a phase later than the one we analyze in detail here (see Fig.~\ref{fig1}d).
RMI is an instability that often happens at the surface between two fluids of different densities, which can be regarded as a special type of RTI \citep{Richtmyer1960,Meshkov1969,Zhou2021rayleigh}. The `traditional' RTI is caused by a constant acceleration (e.g. due to external gravity), and the initial perturbation grows exponentially, whereas the RMI is the result of an impulsive acceleration (e.g. caused by a shock wave), and the initial perturbation grows linearly. Both RTI/RMI instabilities may end up in turbulent dynamics, in which additionally KHI is involved for generation of vortices.

\subsection{Where is the turbulence?}

\begin{figure}
\begin{center}
\includegraphics[width=\linewidth]{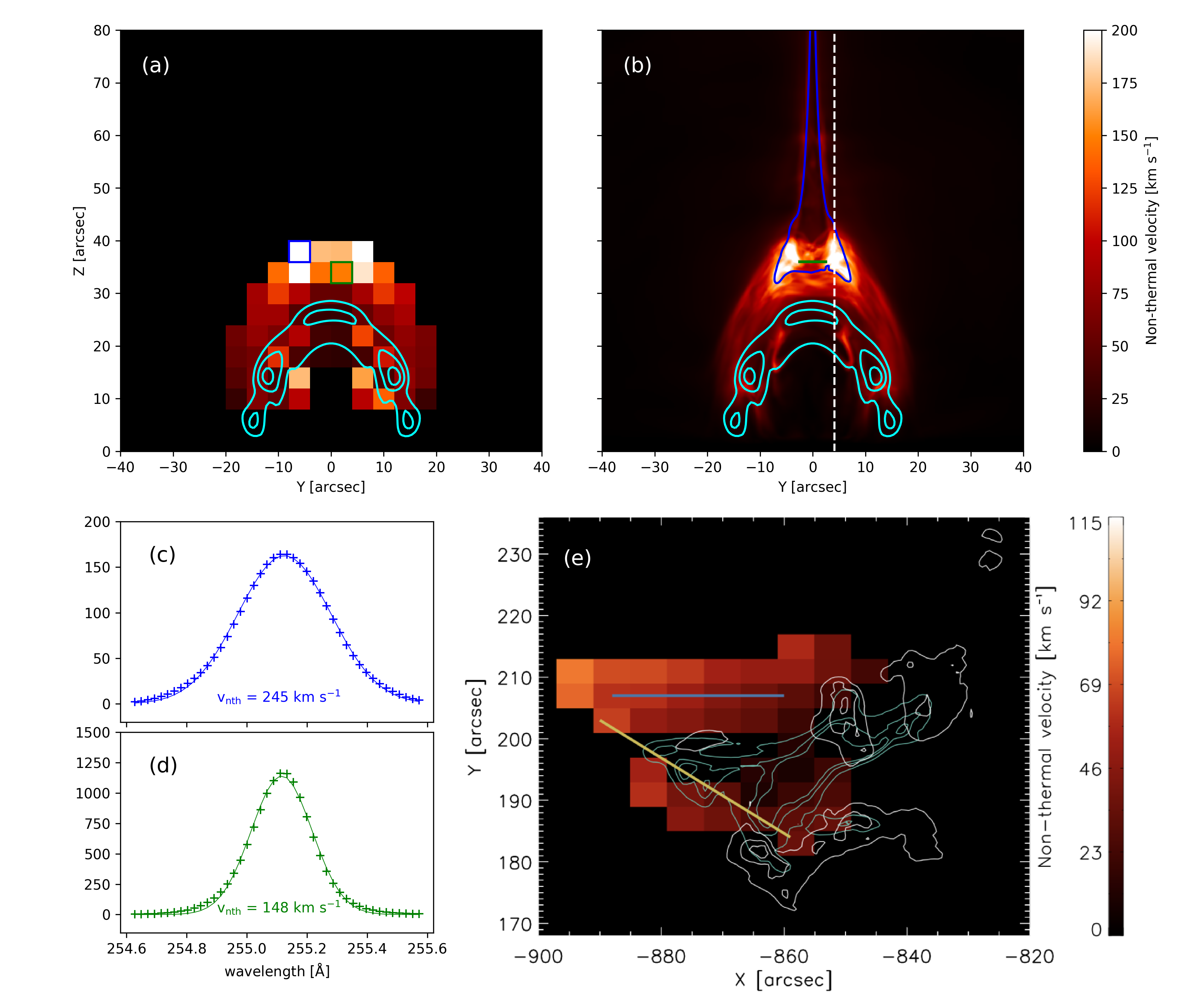}
\caption{(a) Non-thermal velocity distribution obtained from Gaussian fitting of synthesized EIS/255 spectra, where the fitting method refers to \citet{Stores2021}. (b) Non-thermal velocity distribution obtained from plasma density/velocity distribution. Cyan contours in (b) give the locations of the bright loop in Fig.~\ref{fig1}b, where the contour lines show the intensity levels 25\%, 50\% and 80\% of peak 131 {\AA} flux. The regions inside the blue contour have an average magnetic field strength lower than 25 G. The green line gives the approximate location of termination shock (also see Fig.~\ref{fig5}c), but note that the shock locations are different at different X-slices due to interactions between reconnection outflows and magnetic arcades. Panel (c) gives the EIS/255 spectrum inside the blue box in (a) and (d) gives the spectrum inside the green box in (a) (marked with `+'). Solid lines show Gaussian fitting results of the spectra. Panel (e) shows a non-thermal velocity map from the EIS/255 spectra of an observed flare, where the cyan contours give the location of hot/dense flare loop and white contours give the location of footpoints at the solar surface, as taken from \citet{Stores2021}. The spatial resolution of the synthesized EIS/255 data is reduced before fitting, as in \citet{Stores2021}. Panel (a) has a pixel size of 4 arcsec $\times$ 4 arcsec, which is close to that in panel in (e) (6 arcsec $\times$ 4 arcsec). Panel (b) has a smaller pixel size of 0.138 arcsec $\times$ 0.138 arcsec. An animation of this figure is available, showing the evolution of spatial average electron-number density in $x$-direction (panel a) and non-thermal velocity distribution obtained from plasma density/velocity distribution (panel b). The contours in the panels show the intensity levels 10\%, 25\%, 50\% and 80\% of peak 131 {\AA} flux. The animation covers 2.37 minutes of physical time starting at t = 7.51 minutes ( real-time duration 3 s). }
\label{fig2}
\end{center}
\end{figure}

We obtain the turbulent, non-thermal velocities from synthesized spectral profiles of the emission line at 255.1136 {\AA}, corresponding to a peak temperature 10$^{7.2}$ K, as in \citet{Stores2021}. The non-thermal velocity distribution derived from the spectral profiles is given in Fig.~\ref{fig2}a, where the observational result from \citet{Stores2021} is also replicated in Fig.~\ref{fig2}e. The observational data were obtained with the \textit{EUV Imaging Spectrometer} (EIS) onboard \textit{Hinode} \citep{Culhane2007}. The synthesized spectral data adopted the same spatial and wavelength resolutions as the observational data, and the instrument effect has also been included. In order to obtain a clear view on the distribution of turbulence relative to the high density flare loop, we employ a LOS parallel to our $x$ direction (i.e. perpendicular to our flare loops) when synthesizing these EIS/255 spectra. Figure~\ref{fig2}a gives the non-thermal velocity distribution across the $y-z$ plane obtained from the EIS spectra, while a higher resolution equivalent calculated from the available magnetohydrodynamic (MHD) plasma parameters (electron number density $N_{\mathrm{e}}$ and $x$-component of velocity $v_{x}$) by means of
\begin{equation}
v_{\mathrm{nth}} (y,z) = \sqrt{ \frac{ \int N_{\mathrm{e}} (x,y,z) v_{x}^2 (x,y,z) \mathrm{d}x -  [\int N_{\mathrm{e}} (x,y,z) v_{x} (x,y,z) \mathrm{d}x]^2 / \int N_{\mathrm{e}} (x,y,z) \mathrm{d}x }{\int N_{\mathrm{e}} (x,y,z) \mathrm{d}x} } \,,
\label{vnth}
\end{equation}
is shown in panel b. The 131 {\AA} contours indicate the location of the flare EUV loops.

The non-thermal velocity distribution from panels a/b shows very similar features to the observational distribution in panel e, where the turbulence has a wide spatial distribution above the AIA/131 flare loops and the maximum non-thermal velocity is located above the looptop. The maximum non-thermal velocity is located at a weak magnetic field region (see panel b), which is a promising electron acceleration site suggested by \citet{Chen2020} and observationally demonstrated in \citet{Fleishman2022}. The spectral profiles of the high non-thermal velocity regions have perfect Gaussian distributions as demonstrated in Fig.~\ref{fig2} panels c and d, which is a distinctive feature of flare turbulence. 
We confirm from the high resolution non-thermal velocity map (panel b) that there are two isolated high $v_{\mathrm{nth}}$ regions above the high density flare loop, which also show up in panel a. The high non-thermal velocity below the loops in panel a is mainly caused by chromospheric evaporations related to shocks rather than by turbulence.

\subsection{How is the turbulence produced?}

\begin{figure}
\begin{center}
\includegraphics[width=\textwidth]{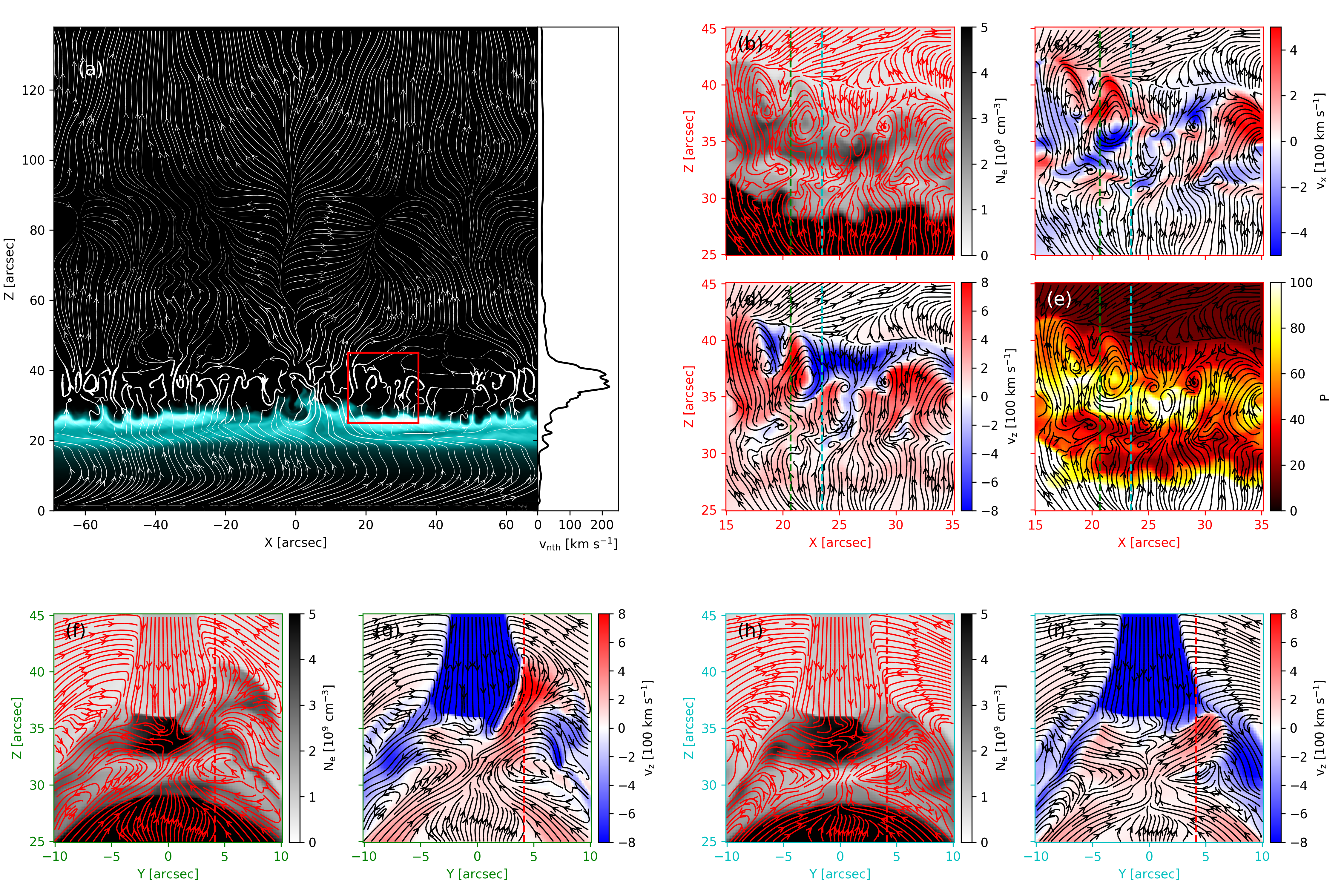}
\caption{Color map in top left panel (a): AIA/131 emission flux distribution at slice $\mathrm{Y}\sim 4$ arcsec, along the white dashed vertical line in Fig.~\ref{fig2}b. At the RHS of panel(a) we show non-thermal velocity. Top right panels (b)-(e): $N_{\mathrm{e}}$, $v_{{x}}$, $v_{{z}}$ and thermal pressure inside the red box in panel (a). Bottom left panels (f)\&(g): $N_{\mathrm{e}}$ and $v_{{z}}$ at a slice marked with a green dashed line in (b)-(e). Bottom right panels (h)\&(i): $N_{\mathrm{e}}$ and $v_{{z}}$ at a slice marked with the cyan dashed line in (b)-(e). The red vertical dashed lines in all bottom panels (f)-(i) give the location of the slice shown in top right panels (b)-(e). Streamlines in (a)-(e) show the velocity field given by $v_x \vec{e}_x + v_z \vec{e}_z$, while that in (f)-(i) show the field given by $v_y \vec{e}_y + v_z \vec{e}_z$. The width of stream lines in panel (a) is in proportion to $\sqrt{v_x^2+v_z^2}$ to highlight the turbulent regions, while that in other panels is constant.}
\label{fig3}
\end{center}
\end{figure}

In order to investigate how the turbulence is produced, we analyze the dynamics in a slice (parallel to the $x$-$z$ plane at $y=3$ Mm or $Y\sim 4$ arcsec) that runs across a region with maximal non-thermal velocities (as indicated by the white dashed vertical line in Fig.~\ref{fig2}b). For that vertical plane, Figure~\ref{fig3}a (top left panel) shows the localized AIA/131 emission flux distribution as well as the `velocity field' (through streamlines of $v_x \vec{e}_x + v_z \vec{e}_z$) on the slice. On its right hand side,  the same panel Figure~\ref{fig3}a gives the non-thermal velocity distribution as function of the height as obtained from Fig.~\ref{fig2}b. It is obvious in the velocity streamline view that there are a lot of vortices around the height with maximal non-thermal velocity (i.e. at $Z \approx 35$ arcsec). 

Figure~\ref{fig3}b-e (four top right panels) show in colorscales the spatial distributions of number density $N_{\mathrm{e}}$, $v_x$, $v_z$, and thermal pressure $P$ (the latter in dimensionless unit) of a selected turbulent region: namely in the red box shown in Fig.~\ref{fig3}a. We each time overlay the streamlines as well. Vortices appear prominently near $Z\sim 35$ arcsec, where we find large shear velocities (e.g. note the velocity jump $>1000$ km s$^{-1}$ in $v_z$ from $X\sim 21$ arcsec to $X\sim 23$ arcsec), which indicates the turbulence is produced via KHI. 
KHI may happen at the interface between high speed shear flows to produce vortices at their interface. This instability is often inhibited by magnetic tension in magnetized plasmas. When the flow velocity vector is parallel to the magnetic field lines, triggering KHI requires that (half) the velocity jump is larger than the local Alfv\'en speed \citep{Keppens1999}.
Here, in the slice analysed, we find conditions that are quite different from this standard 2D situation. We have here a guiding magnetic field that is locally almost perpendicular to the $x$-$z$ plane shown here (shape of the guiding field refers to the white solid line in Fig.~\ref{fig4}a), while there are large shears in both $v_x$ and $v_z$. 
As a result, KHI vortices can be produced on this $x$-$z$ plane without distorting the magnetic field too much, and hence without producing a stabilizing magnetic tension. Such a condition is unconditionally unstable for KHI. The relative orientation of magnetic field and flow shear is similar in the case of so-called TWIKH rolls from transverse-wave-induced KHI in coronal loops (e.g. \citealp{Antolin2014,Guo2019,Shi2021}). A simplified growth rate estimate of KHI in such a condition $\omega = k v_0$ (using equation~13.38 in \citealp{Goedbloed2019}) gives a growth timescale ($1/\omega$) of seconds, the same order of magnitude as the vortex generation time scale seen in our simulation. Here, we used $k = 1/(\mathrm{several\ Mm})$ since our vortices have a length scale of Mm and a half shear speed $v_0$ of hundreds up to one thousand km s$^{-1}$, as found in the flow distribution.

The initial shear motions come from non-linear interactions observed in the downward reconnection flows with the magnetic arcades above the AIA/131 loop. Collisions between the downflows and the arcades leads to the formation of a termination shock and complicated reflected upflows. Their spatial distribution on the vertical $y$-$z$ plane is different for different $x$ (as shown in the bottom panels Fig.~\ref{fig3}g \& i), which enhances shear motion between downward flows and upward reflection flows. Combining all views collected in Fig.~\ref{fig3} we see that vortices prefer to form on $x$-$z$ planes, which is perpendicular to the local magnetic field direction. A related, detailed study on the interaction between reconnection flows and their arcades in 2D conditions is found in \citet{Ye2021}. 
The magnetic field strength inside the flaring region is spatial-dependent, where the legs of the guiding line shown in Fig.~\ref{fig4}a have an average strength of $\sim 40$ G, while the apex of the line has a weaker strength of less than 20 G due to the impact of turbulent motion. In real flares, the magnetic field can be much stronger, with field strength higher than 100 G above the 131 {\AA} loops. However, turbulence can still be produced with the mechanism we mentioned above, as it is difficult for the magnetic field to inhibit the growth of KHI when the flows velocity vectors are perpendicular to the field lines.

Note further that the turbulent upper border of the AIA/131 loops in Fig.~\ref{fig3}a is attributed by  \citet{Shen2022} to RTI/RMI effects. These authors focused on the dark fingers as also seen in an as yet early stage in our Fig.~\ref{fig3}a, where the connection with Supra-Arcade Downflows was made by \citet{Shen2022}. 
The RTI/RMI region does not lead to large turbulent velocities during our simulation and is clearly situated below the region of maximal non-thermal velocities. As demonstrated in Fig.~\ref{fig3}a, the non-thermal velocity around the upper border of the AIA/131 loops is smaller than 50 km s$^{-1}$. Our (higher resolution) simulation demonstrates that the turbulent zone is foremost located above the EUV loops (see also Fig.~\ref{fig2}) and most likely driven by KHI mechanisms.

\subsection{Turbulence at the lower atmosphere}

\begin{figure}
\begin{center}
\includegraphics[width=\textwidth]{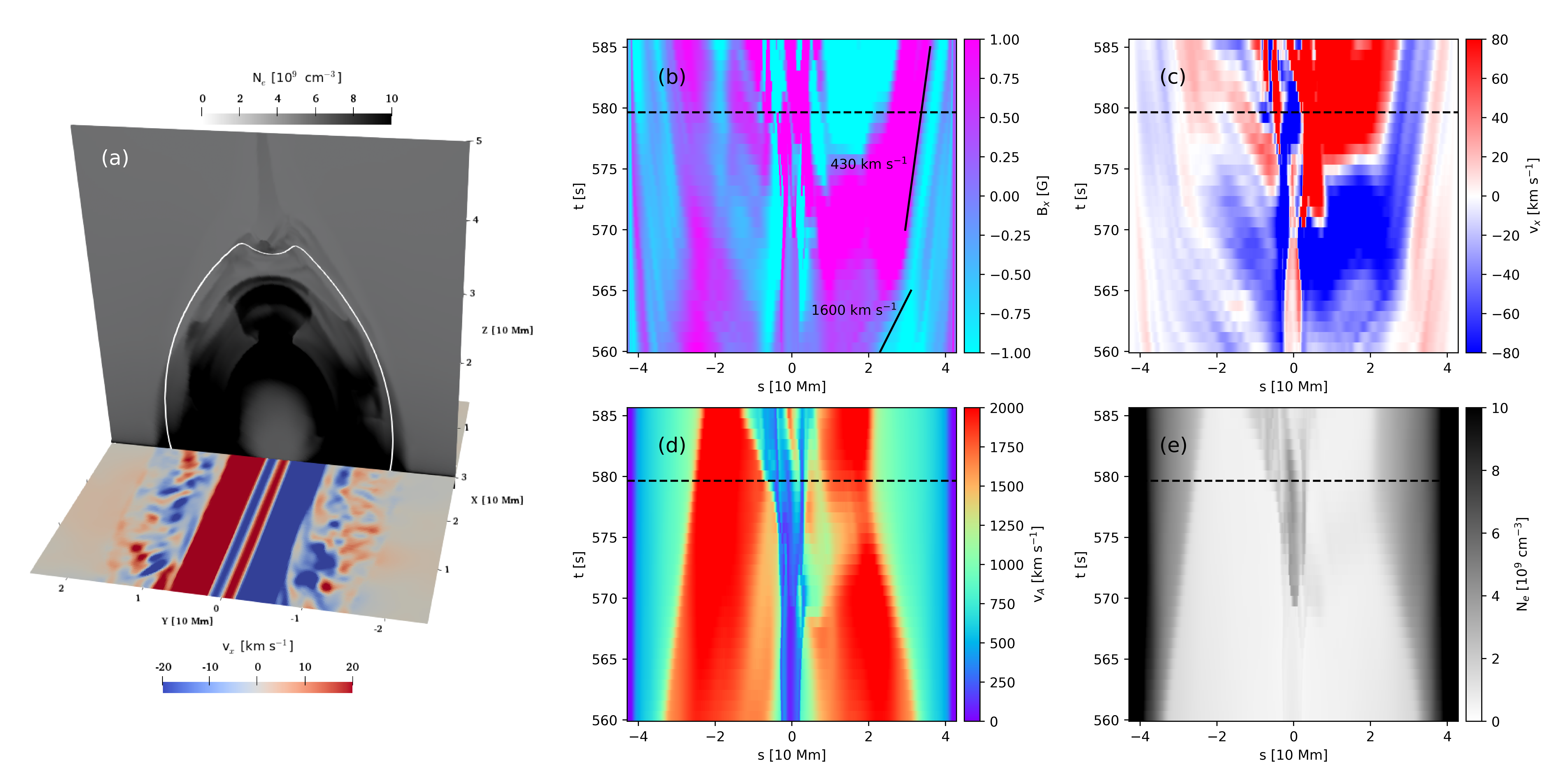}
\caption{Horizontal slice in the left 3D view panel (a): $v_x$ distribution at $z=2$ Mm, while the vertical slice shows the $N_{\mathrm{e}}$ distribution at $x=30$ Mm. Right panels (b)-(e): Time-space plots of $B_x$, $v_x$, Alfv\'en speed $v_{\mathrm{A}}$ and $N_{\mathrm{e}}$ along the white solid (field) line shown in (a), where the minimum $s$-value is located at the left end of this line. The horizontal dashed lines in these panels (b)-(e) and the vertical dotted line in Fig.~\ref{fig1}d give the corresponding time for the 3D view (a) at left.}
\label{fig4}
\end{center}
\end{figure}

Turbulence can also be found at lower atmospheric regions in our simulation. The bottom slice in the 3D view at left in Fig.~\ref{fig4}a demonstrates the $v_x$ distribution at a layer at height $z=2$ Mm. Turbulent motions up to $\sim$ 10 km s$^{-1}$ appear at the footpoints of the high density arcades. These are shown in the vertical slice in the 3D view of Fig.~\ref{fig4}a, in their number density distribution at $x=30$ Mm ($X\sim$ 41 arcsec). The turbulent region at the $z=2$ Mm layer has an average number density of $\sim$10$^{11}$ cm$^{-3}$ and an average temperature of $\sim$1 MK. The temperature actually drops rapidly from several MK to a typical chromospheric temperature of $\sim$0.01 MK around this height. Turbulent motions can still be found at a lower $z=1$ Mm layer where the average number density is $\sim$10$^{13}$ cm$^{-3}$, but with reduced speeds by one order of magnitude compared to those at $z=2$ Mm.

We find that the turbulence at this lower atmosphere region gets propagated from higher regions downwards, rather than being generated locally. We select a (time-invariant) curve which gives a general direction of the magnetic field (the white solid line in Fig.~\ref{fig4}a) and study the time development of typical MHD quantities along that curve. The time-space plots of $B_x$ and $v_x$ along the curve (top right panels) demonstrate that there are structures propagating from the middle of this curve to its ends. The $B_x$ and $v_x$ are seen to show an anti-phase when structures are propagating toward $s>0$, while they are in phase when the structures are propagating toward $s<0$. Since the magnetic field vector is pointing from $s<0$ to $s>0$, such a phase relationship indicates that the structures are Alfv\'enic perturbations. In fact, the propagation speeds (fitted speeds are indicated in Fig.~\ref{fig4}b) of the structures are close to the local Alfv\'en speeds which are quantified in panel (d). Sudden decreases in propagating speed around $s = 30$ Mm are a result of a change in the local Alfv\'en speed, where this change in Alfv\'en speed is caused by plasma density variation that is due to chromospheric evaporation processes, as evident from the time-space plot of number density in panel (e).
The downward propagating Alfv\'enic perturbations and the turbulence in the chromosphere may contribute to the generation of energetic electrons (e.g. \citealp{Fletcher2008}). The Alfv\'enic perturbations may bring a lot of energy to the lower atmosphere, and contribute to the heating of the chromosphere and the generation of evaporations (e.g. \citealp{Russell2013,Reep2016}). Those effects are worth investigating in future studies.

The bigger picture obtained from all previous sections is as follows: turbulence gets produced above the looptops due to KHI, and this turbulence gets propagated along magnetic fields to lower regions, leading to a spatially distributed turbulent plasma at all heights. The arcade-shape spatial distribution of the non-thermal velocity regions seen in Fig.~\ref{fig2}a and b also supports this bigger picture, in line with the arcade-shape magnetic field that forms during the flare process.

\subsection{Turbulent Kinetic Energy}

When connecting MHD turbulence with actual non-thermal particle acceleration, we should still verify whether the released magnetic energy from reconnection can be efficiently converted to turbulence energy. For KHI driven turbulence, the energy in the turbulence comes from the kinetic energy of bulk flows, the reconnection downflows in particular. Therefore, here we compare the time-evolving turbulence energy with the time-integrated reconnection downflow kinetic energy. 

\begin{figure}
\begin{center}
\includegraphics[width=\textwidth]{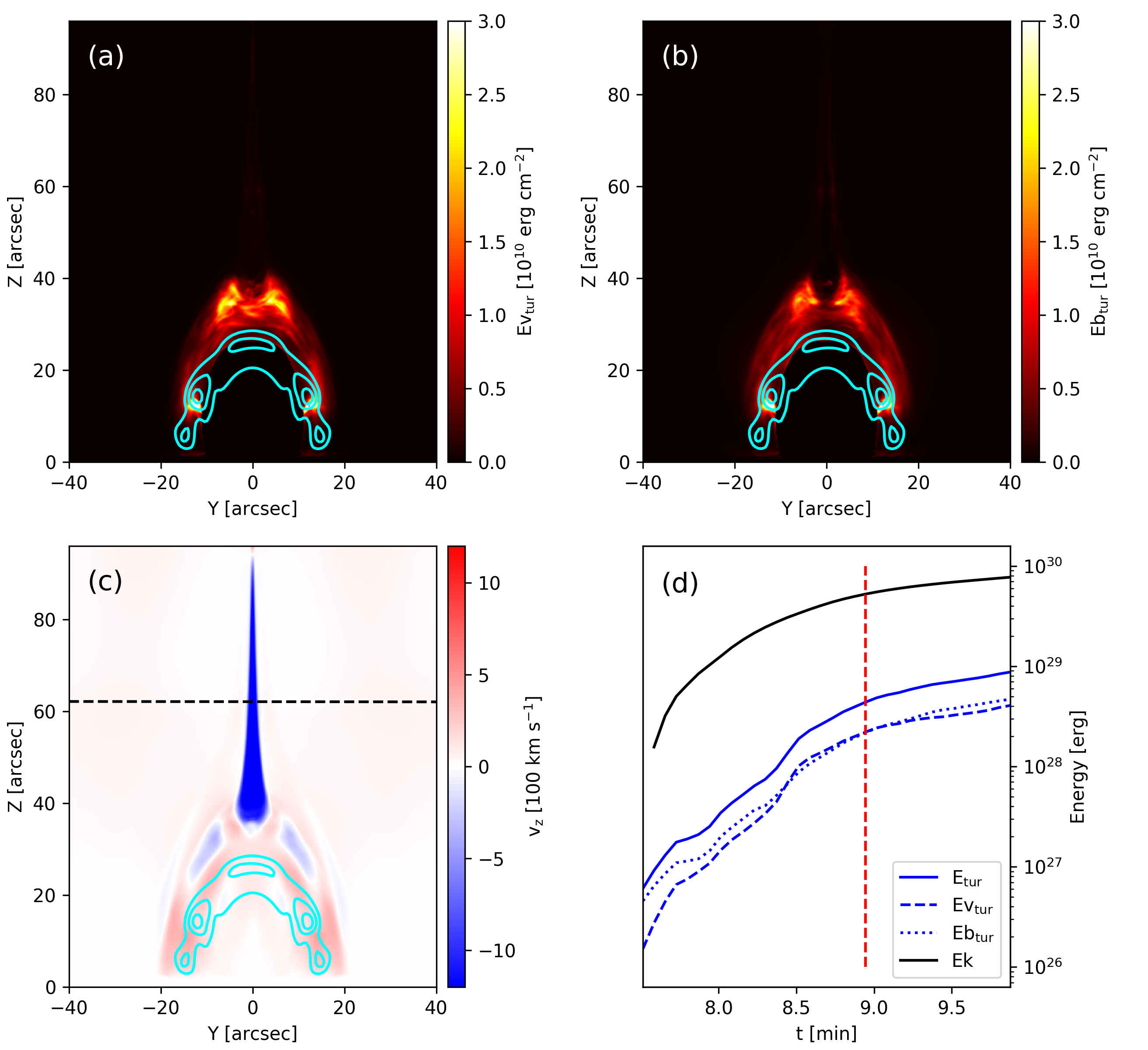}
\caption{(a) Spatial distribution of turbulent kinetic energy. (b) Spatial distribution of turbulent magnetic energy. (c) X-averaged vertical speed distribution. Taken along the black solid line from this panel (c), we show in (d) the time-integrated reconnection downflow kinetic energy (black solid line) that goes through the surface. Panel (d) further shows in blue solid line the instantaneous total turbulence energy, and its division over kinetic energy and turbulent magnetic energy. In this panel (d), the vertical red solid line corresponds to the time for panels (a), (b) and (c). As in Fig.~\ref{fig2}, the contours in panels (a),(b) and (c) show the AIA/131 loop location.}
\label{fig5}
\end{center}
\end{figure}

Turbulence energy consist of kinetic energy of the turbulent velocity field and magnetic energy of the turbulent magnetic field. The spatial distribution of the kinetic energy density and the magnetic energy density at $t \approx 9$ min is shown in Fig.~\ref{fig5}a and b, respectively. The turbulent kinetic energy density is calculated from
\begin{equation}
Ev_{\mathrm{tur}} (y,z) = \int \rho (x,y,z) v_{x}^2 (x,y,z) \mathrm{d}x -  [\int \rho (x,y,z) v_{x} (x,y,z) \mathrm{d}x]^2 / \int \rho (x,y,z) \mathrm{d}x ,
\end{equation}
and the turbulent magnetic energy density is calculated from
\begin{equation}
Eb_{\mathrm{tur}} (y,z) = \int  B_{x}^2 (x,y,z) \mathrm{d}x -  [\int  B_{x} (x,y,z) \mathrm{d}x]^2 / \int \mathrm{d}x,
\end{equation}
where $\rho$ is density. An assumption applied in the calculation is that the turbulence is anisotropic (Alfv\'enic), such that the turbulent motion is (only) freely developed in directions perpendicular to the (average) magnetic field, which is why we only incorporate $x$-components in both flow and magnetic field turbulent quantifications for this $y-z$ view. Fig.~\ref{fig5}a and b demonstrate that $Ev_{\mathrm{tur}}$ and $Eb_{\mathrm{tur}}$ have similar density and spatial distribution, again supporting our assumption that the turbulence is indeed Alfv\'enic. The time-integrated kinetic energy of the reconnection downflow is calculated via integrating the kinetic energy flux that goes across a slice at height $z=45$ Mm ($Z\sim 62$ arcsec, the black dashed line in Fig.~\ref{fig5}c), where we do the integration from $t=450$ s, the time where the fast reconnection regime starts.
The color map shows the spatial average vertical velocity distribution given by
\begin{equation}
\bar{v}_z (y,z) = \int N_{\mathrm{e}} (x,y,z) v_{z} (x,y,z) \mathrm{d}x / \int N_{\mathrm{e}} (x,y,z) \mathrm{d}x,
\end{equation}
which gives locations of the reconnection downflow.

Figure~\ref{fig5}d gives a comparison of the (time) integrated reconnection downflow kinetic energy $Ek$, the instantaneous total turbulence energy $E_{\mathrm{tur}}$, and the kinetic and magnetic components of this turbulence energy ($E_{\mathrm{tur}}=$$Ev_{\mathrm{tur}}$ $+$ $Eb_{\mathrm{tur}}$). $Ev_{\mathrm{tur}}$ closely follows $Eb_{\mathrm{tur}}$ during the entire time evolution as expected, a feature of Alfv\'enic motions. $E_{\mathrm{tur}}$ shows a similar tendency as $Ek$, where we find $E_{\mathrm{tur}} \approx 0.1 Ek$, so roughly a 10 \% fraction at each time. The total, time-integrated amount that is transferred from bulk kinetic energy of the reconnection downward flows to turbulence energy is more than the instantaneous turbulence energy $Ek$, considering that turbulence energy is continuously (transported to the chromosphere and) diffused.

\section{Conclusions} \label{sec:conclu}

The spatial distribution and the characteristic velocity of the plasma turbulence in flare observation are successfully reproduced in our 3D MHD flare simulation. 
The maximum turbulent velocity reached is higher than 200 km s$^{-1}$, and is located in close connection to the reconnection termination shock. Reconnection downflows collide with magnetic arcades near the termination shock and then produce complicated reflection upflows. Non-linear interactions between the downflows and upflows lead to spatially extended regions where high speed turbulent motion forms naturally, in which KHI dominates. More than 10\% of the downflow kinetic energy is converted into turbulence energy via this mechanism. The turbulence is anisotropic due to magnetic tension, as the turbulent vortices only appear most clearly inside the plane that is perpendicular to the guiding magnetic field. The locally generated turbulence can propagate along the guiding magnetic field as Alfv\'enic perturbations, and then leads to a wider spatial distribution of turbulent motions. The lower atmosphere, where the number density reaches $10^{11}$ cm$^{-3}$, still has turbulent motions of order 10 km s$^{-1}$. 

Our model works for any flare that involves magnetic reconnection, no matter how big the flare is or how long it takes to release energy, since it is governed by the scale-invariant MHD description. Downwards reconnection outflows invariably encounter lower-lying magnetic fields, by which the outflows are stopped and diverted in the direction perpendicular to the magnetic field, making the flow shear orthogonal to the field. Such a condition is a perfect environment for KHI in full 3D (and missed in 2D settings), regardless of the magnetic field strength. KHI can be easily triggered for various reasons in such a condition, where the interaction between the downflows and reflected upflows is (only) one of them.

Our simulations used various algorithmic improvements that render it possible to resolve details down to few 100 km, thanks to adaptive mesh refinement, and the TRAC treatment to properly evolve multi-dimensional MHD from chromosphere to corona. Application of this TRAC method is not crucial for this KHI turbulence at the studied impulsive phase, but will be important for properly generating coronal rain in the gradual phase~\citep{Ruan2021}. A 3D numerical study on the formation of post-flare coronal rain can serve as interesting follow-up to this work. Then, the role of other instabilities (like RTI/RMI and thermal instabilities) in various evolutionary stages of a 3D flare loop can be clarified in detail. 
The maximum non-thermal velocity is located at a weak magnetic field region, where the high-energy electrons are accelerated. It indicates that plasma turbulence probably plays an essential role in accelerating flare electrons. 
The next step of this study should address the consequences of the fluid turbulence for charged particle dynamics more seriously, either through test particle assessment of the attainable particle acceleration efficiency, or even evolving to hybrid (fluid/particle) or fully kinetic models. There is no doubt that turbulence can accelerate electrons/ions to high energies, but how much of the energy released by reconnection can go to energetic electrons/ions in this way is still a question. According to \citet{Aschwanden2017}, about half of the energy will go to the energetic electrons and about 17\% of the energy will go to the energetic ions on average. \citet{Emslie2012} quotes lower acceleration efficiency, but still about 20\% of the released energy would go to the energetic electrons and ions. It might be difficult to achieve this in a multi-step process such as turbulence or shock acceleration \citep{Cargill1996,Miller1997}. Most likely, the tens of percent energy transfer efficiency is the result of multiple acceleration mechanisms cooperating (e.g., turbulence acceleration and shock acceleration). Note that other scenarios can achieve high acceleration efficiencies, such as the acceleration at fragmented current sheets \citep{Cargill2012}.

\begin{acknowledgments}
We thank the referee Peter Cargill for very constructive comments.
WR was supported by a postdoctoral mandate (PDMT1/21/027) by KU Leuven.
LY was supported by the Youth Innovation Promotion Association of CAS (2021064). RK is supported by Internal funds KU Leuven through 
the project C14/19/089 \nobreak{TRACESpace,} and an FWO project G0B4521N. RK also received funding from the European Research 
Council (ERC) under the European Union Horizon 2020 research and innovation programme (grant agreement No. 833251 \nobreak{PROMINENT} ERC-ADG 2018).
\end{acknowledgments}

\bibliography{Ruan2022}{}
\bibliographystyle{aasjournal}

\end{document}